\documentclass[sigconf, screen]{acmart}

\AtBeginDocument{%
  \providecommand\BibTeX{{%
    \normalfont B\kern-0.5em{\scshape i\kern-0.25em b}\kern-0.8em\TeX}}}

\setcopyright{acmcopyright}

\copyrightyear{2026}
\acmYear{2026}
\setcopyright{cc}
\setcctype{by}
\acmConference[CHI '26 Workshop]{Workshop on Restoring Human Authenticity in AI-Mediated Communication at CHI '26}{April 13--17, 2026}{Barcelona, Spain}
\acmBooktitle{Workshop on Restoring Human Authenticity in AI-Mediated Communication (AI-MC) at the 2026 CHI Conference on Human Factors in Computing Systems (CHI '26), April 13--17, 2026, Barcelona, Spain}
\acmPrice{}
\acmDOI{XX.XXXX/XXXXXXX.XXXXXXX}
\acmISBN{XXX-X-XXXX-XXXX-X/2026/04}

\begin{document}

\title{Rethinking AI-Mediated Minority Support in Power-Imbalanced Group Decision-Making: From Anonymity To Authenticity}

\author{Soohwan Lee}
\orcid{0000-0001-8652-3408}
\affiliation{%
  \institution{Department of Design, UNIST}
  \city{Ulsan}
  \country{Republic of Korea}}
\email{soohwanlee@unist.ac.kr}

\author{Kyungho Lee}
\orcid{0000-0002-1292-3422}
\affiliation{\institution{Department of Design, UNIST}
\city{Ulsan}
\country{Republic of Korea}}
\email{kyungho@unist.ac.kr}

\begin{abstract}
AI-mediated Communication (AIMC) systems increasingly aim to protect minority voices by anonymizing or proxying their input, but anonymity and authenticity are not the same construct. This position paper draws on an ongoing empirical study comparing two LLM-powered minority support strategies in hierarchical group decision-making. We found that relaying minority input anonymously through AI increased participation but significantly reduced psychological safety and satisfaction, while generating only autonomous counterarguments improved satisfaction and reduced marginalization. These counterintuitive findings reveal three provocations for AIMC design in hierarchical contexts: the inherent trade-offs among anonymity, authenticity, agency, and accountability; the risk that power asymmetry reverses intended effects; and the need for AI to facilitate group reflection rather than substitute for human responsibility. These findings and provocations are offered as a contribution to the \textit{Restoring Human Authenticity in AI-Mediated Communication} workshop.
\end{abstract}



\maketitle

\section{Motivation and Background}
Power imbalances in group decision-making frequently suppress minority perspectives and weaken collective outcomes \cite{janisGroupthinkPsychologicalStudies1982}. Compliance pressures lead individuals to publicly align with the majority despite private disagreement, undermining psychological safety and discouraging participation from less-empowered members \cite{kelmanComplianceIdentificationInternalization1958, moscoviciStudiesSocialInfluence1976}. Recent HCI research has introduced AI-powered interventions to mitigate groupthink and encourage broader participation \cite{chiangEnhancingAIAssistedGroup2024}, but many assume relatively equal standing among participants. Comparatively less explored is how AI can represent and route dissent when dissenters face social risk from identification or retaliation in power-imbalanced situations \cite{hwangSoundSupportGendered2024, liImprovingNonNativeSpeakers2022, leeConversationalAgentsCatalysts2025b}.

This position paper draws on an ongoing empirical study comparing two LLM-powered minority support strategies in hierarchical group decision-making. We found that anonymously relaying minority input through AI increased participation but significantly reduced psychological safety and satisfaction, while generating only autonomous counterarguments improved satisfaction and reduced marginalization. These counterintuitive findings suggest that in hierarchical settings, how AI introduces dissent and whose voice it is perceived to carry matters as much as whether dissent enters the discussion at all. This paper contributes to \textit{CHI'26 Workshop on Restoring Human Authenticity in AI-Mediated Communication} by arguing that anonymity and authenticity are distinct properties that current AIMC designs tend to conflate. We summarize our empirical findings, discuss how power asymmetry complicates assumptions in existing AIMC work, and offer design provocations for preserving expressive ownership in hierarchical group settings.

\begin{figure*}[h]
  \centering
  \includegraphics[width=0.85\textwidth]{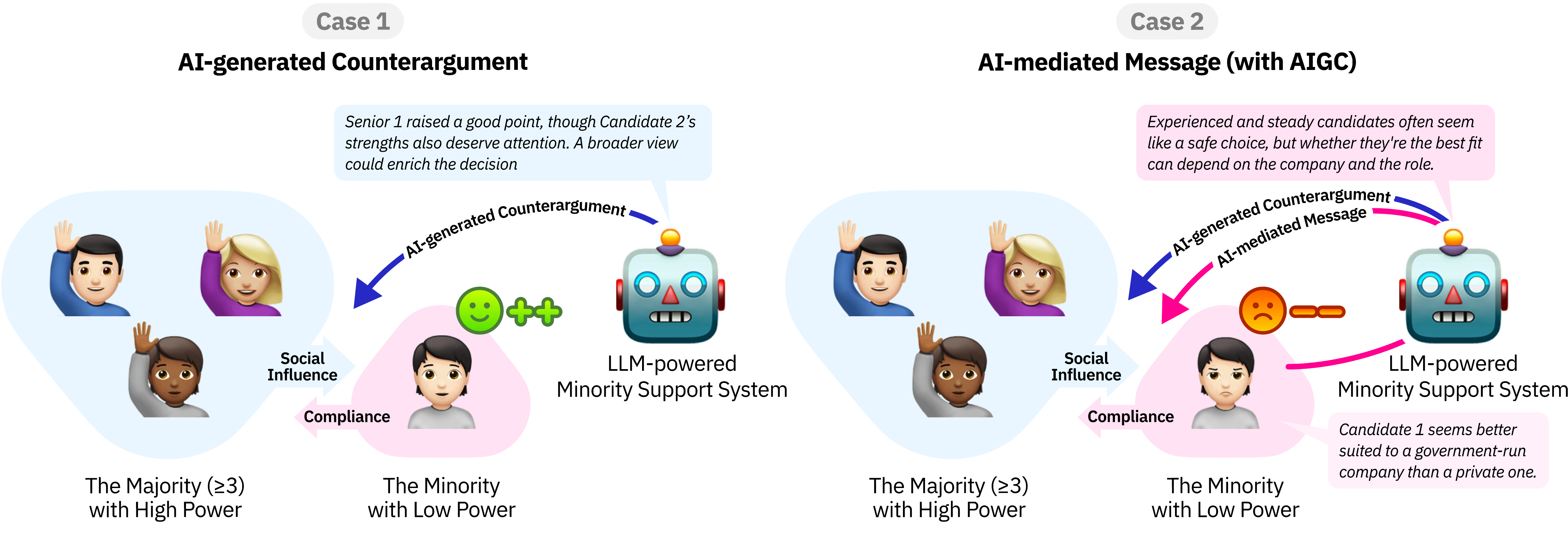}
  \caption{An LLM-powered minority support system mediates majority-minority dynamics through two designs: AIGC, which autonomously posts counterpoints to normalize dissent, and AIMM, which paraphrases privately submitted minority views blended with AIGC outputs to obscure authorship. AIGC improved group atmosphere and satisfaction, whereas AIMM increased participation but paradoxically undermined minority members' psychological safety and satisfaction.}
  \Description{This figure illustrates four patterns of AI-mediated communication in group settings, represented by diagrams labeled A–D with arrows indicating information flow. (A) The human requests AI-generated content and relays it to another participant. (B) The human selectively shares AI output with others. (C) The AI reformulates a human’s message and presents it directly. (D) The AI autonomously facilitates communication between participants. Numbered arrows indicate the sequence of interactions in each pattern, showing different levels of AI involvement in mediating group discussions.}
  \label{fig:teaser}
\end{figure*}

\begin{figure*}[h]
  \centering
  \includegraphics[width=0.8\textwidth]{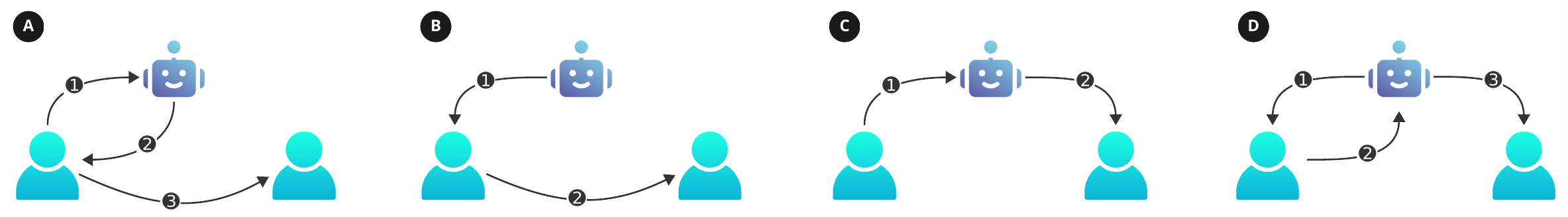}
  \caption{Four patterns of AI-mediated group communication\cite{leeAmplifyingMinorityVoices2025b}: (A) a human relays AI-generated content to the group; (B) a human selectively discloses AI output to participants; (C) the AI reformulates a human's message before presenting it to the group; and (D) the AI directly facilitates inter-participant communication. Arrows denote information flow, with numbered sequences indicating interaction order.}
  \Description{This figure illustrates four patterns of AI-mediated communication in group settings, represented by diagrams labeled A–D with arrows indicating information flow. (A) The human requests AI-generated content and relays it to another participant. (B) The human selectively shares AI output with others. (C) The AI reformulates a human’s message and presents it directly. (D) The AI autonomously facilitates communication between participants. Numbered arrows indicate the sequence of interactions in each pattern, showing different levels of AI involvement in mediating group discussions.}
  \label{fig:communicationPattern}
\end{figure*}

\section{Summary of Empirical Findings}
We conducted a mixed-methods experiment with 96 Korean participants in 24 four-person groups, each consisting of three high-power majority members (seniors) and one low-power minority member (junior). Groups completed two corporate decision-making tasks under one of three conditions (\autoref{fig:teaser}): a no-AI baseline, an AI-generated Counterargument condition (AIGC), or an AI-mediated Message condition (AIMM). In AIGC, the system autonomously posts counterpoints to the emerging majority consensus. In AIMM, the system also accepts private input from the minority member, paraphrases it, and relays it (\autoref{fig:communicationPattern}-C) as its own output alongside its own counterarguments, making the two message types indistinguishable in terms of provenance. We measured psychological safety, marginalization, engagement, satisfaction, and cognitive workload via validated self-report scales and behavioral metrics, supplemented by semi-structured exit interviews.

Final group decisions aligned with senior preferences in 80\% of cases across all conditions, suggesting that neither intervention was sufficient to shift collective outcomes. AIGC reduced minority members' marginalization and improved their satisfaction without negatively affecting psychological safety, as autonomous counterarguments fostered a more open group atmosphere in which minority members felt less isolated in their dissent. In contrast, AIMM increased participation volume but significantly reduced psychological safety and satisfaction. Majority members discounted AI-authored messages because they perceived them as non-human, leaving minority members feeling ignored even when their views were relayed. This produced invisibility — authorship was hidden, and the message was devalued as AI-generated.

\section{Future Design Provocations}
Our findings suggest three tensions that AIMC research and design should address more directly in hierarchical group contexts.

\textbf{Anonymity, Authenticity, Accountability, and Agency Involve Inherent Trade-offs.} Anonymity is often discussed as one mechanism for protecting minority voices \cite{stauferSilencingRiskNot2024}, but our results suggest it is neither sufficient nor straightforwardly beneficial in hierarchical settings. Full anonymity allows minority members to express dissent without social risk, but can simultaneously strip them of expressive ownership, visibility, and the sense that their contribution is genuinely theirs. The inverse is equally problematic: speaking directly preserves authenticity, agency, and accountability, but exposes the speaker to marginalization and psychological distress. Designing for authenticity in AIMC therefore requires attending not only to anonymity, but to how systems can support minority members in feeling genuinely heard, credited, and responsible for their own contributions.

\textbf{Power Asymmetry Can Reverse the Intended Effects of AIMC.} Most AIMC systems are designed and evaluated under conditions of relatively equal participation. In hierarchical settings, however, AI-authored messages may be systematically discounted by dominant group members, who may actively rebut AI interventions to reinforce their own position rather than engaging with the underlying dissent. What is designed to foster inclusion can, in practice, deepen existing power imbalances. Future AIMC research should treat power asymmetry as a primary design variable rather than a contextual footnote.

\textbf{AI Should Prompt Reflection, Not Replace Organizational Responsibility.} The organizational contexts where such systems are most needed may also be least likely to adopt them, or may adopt them in ways that serve majority rather than minority interests. Effective deployment therefore requires examining not only system efficacy but also the adoption incentives of senior stakeholders and the motivations of members who will use the system. More broadly, future systems might more sustainably foster inclusive atmospheres by prompting collective reflection and encouraging groups to find better directions on their own. Ultimately, the goal should be organizations that no longer need such systems because psychological safety \cite{edmondsonPsychologicalSafetyLearning1999} is already embedded in their culture.

\section{Conclusion}
This position paper argues that anonymity and authenticity are distinct constructs that AIMC design must address separately. Our empirical findings show that AI-mediated anonymity can increase participation while simultaneously eroding expressive ownership and psychological safety. Future AIMC systems should move beyond protecting the volume of minority speech toward ensuring that minority voices are genuinely heard, attributed, and acknowledged within their group's relational dynamics.

\begin{acks}
This research was partially supported by a grant from the Korea Institute for Advancement of Technology (KIAT) funded by the Government of Korea (MOTIE) (P0025495, Establishment of Infrastructure for Integrated Utilization of Design Industry Data).
\end{acks}

\bibliographystyle{ACM-Reference-Format}
\bibliography{devilsAdvocate}

@inproceedings{leeAmplifyingMinorityVoices2025b,
author = {Lee, SooHwan and Kim, Mingyu and Hwang, Seoyeong and Kim, Dajung and Lee, Kyungho},
title = {Amplifying Minority Voices: AI-Mediated Devil's Advocate System for Inclusive Group Decision-Making},
year = {2025},
isbn = {9798400714092},
publisher = {Association for Computing Machinery},
address = {New York, NY, USA},
url = {https://doi.org/10.1145/3708557.3716334},
doi = {10.1145/3708557.3716334},
booktitle = {Companion Proceedings of the 30th International Conference on Intelligent User Interfaces},
pages = {17–21},
numpages = {5},
location = {
},
series = {IUI '25 Companion}
}

@inproceedings{leeConversationalAgentsCatalysts2025b,
  title = {Conversational {{Agents}} as {{Catalysts}} for {{Critical Thinking}}: {{Challenging Social Influence}} in {{Group Decision-making}}},
  shorttitle = {Conversational {{Agents}} as {{Catalysts}} for {{Critical Thinking}}},
  booktitle = {Proceedings of the {{Extended Abstracts}} of the {{CHI Conference}} on {{Human Factors}} in {{Computing Systems}}},
  author = {Lee, Soohwan and Hwang, Seoyeong and Kim, Dajung and Lee, Kyungho},
  year = 2025,
  month = apr,
  series = {{{CHI EA}} '25},
  pages = {1--12},
  publisher = {Association for Computing Machinery},
  address = {New York, NY, USA},
  doi = {10.1145/3706599.3719792},
  urldate = {2025-12-18},
  isbn = {979-8-4007-1395-8}
}

@inproceedings{chiangEnhancingAIAssistedGroup2024,
  title = {Enhancing {{AI-Assisted Group Decision Making}} through {{LLM-Powered Devil}}'s {{Advocate}}},
  booktitle = {Proceedings of the 29th {{International Conference}} on {{Intelligent User Interfaces}}},
  author = {Chiang, Chun-Wei and Lu, Zhuoran and Li, Zhuoyan and Yin, Ming},
  year = {2024},
  month = apr,
  series = {{{IUI}} '24},
  pages = {103--119},
  publisher = {Association for Computing Machinery},
  address = {New York, NY, USA},
  doi = {10.1145/3640543.3645199},
  urldate = {2024-04-16},
  isbn = {979-8-4007-0508-3}
}

@article{edmondsonPsychologicalSafetyLearning1999,
  title = {Psychological {{Safety}} and {{Learning Behavior}} in {{Work Teams}}},
  author = {Edmondson, Amy},
  year = {1999},
  month = jun,
  journal = {Administrative Science Quarterly},
  volume = {44},
  number = {2},
  pages = {350--383},
  publisher = {SAGE Publications Inc},
  issn = {0001-8392},
  doi = {10.2307/2666999},
  urldate = {2024-08-16},
  langid = {english}
}

@inproceedings{hwangSoundSupportGendered2024,
  title = {The {{Sound}} of {{Support}}: {{Gendered Voice Agent}} as {{Support}} to {{Minority Teammates}} in {{Gender-Imbalanced Team}}},
  shorttitle = {The {{Sound}} of {{Support}}},
  booktitle = {Proceedings of the {{CHI Conference}} on {{Human Factors}} in {{Computing Systems}}},
  author = {Hwang, Angel Hsing-Chi and Won, Andrea Stevenson},
  year = {2024},
  month = may,
  series = {{{CHI}} '24},
  pages = {1--22},
  publisher = {Association for Computing Machinery},
  address = {New York, NY, USA},
  doi = {10.1145/3613904.3642202},
  urldate = {2024-07-16},
  isbn = {979-8-4007-0330-0}
}

@book{janisGroupthinkPsychologicalStudies1982,
  title = {Groupthink : Psychological Studies of Policy Decisions and Fiascoes},
  shorttitle = {Groupthink},
  author = {Janis, Irving L. (Irving Lester)},
  year = {1982},
  publisher = {Boston : Houghton Mifflin},
  urldate = {2024-08-09},
  collaborator = {{Internet Archive}},
  isbn = {978-0-395-31704-4},
  langid = {english}
}

@article{kelmanComplianceIdentificationInternalization1958,
  title = {Compliance, Identification, and Internalization Three Processes of Attitude Change},
  author = {Kelman, Herbert C.},
  year = {1958},
  month = mar,
  journal = {Journal of Conflict Resolution},
  volume = {2},
  number = {1},
  pages = {51--60},
  publisher = {SAGE Publications Inc},
  issn = {0022-0027},
  doi = {10.1177/002200275800200106},
  urldate = {2024-07-18},
  langid = {english}
}

@article{liImprovingNonNativeSpeakers2022,
  title = {Improving {{Non-Native Speakers}}' {{Participation}} with an {{Automatic Agent}} in {{Multilingual Groups}}},
  author = {Li, Xiaoyan and Yamashita, Naomi and Duan, Wen and Shirai, Yoshinari and Fussell, Susan R.},
  year = {2022},
  month = dec,
  journal = {Proc. ACM Hum.-Comput. Interact.},
  volume = {7},
  number = {GROUP},
  pages = {12:1--12:28},
  doi = {10.1145/3567562},
  urldate = {2024-07-19}
}

@article{moscoviciStudiesSocialInfluence1976,
  title = {Studies in Social Influence {{III}}: {{Majority}} versus Minority Influence in a Group},
  shorttitle = {Studies in Social Influence {{III}}},
  author = {Moscovici, Serge and Lage, Elisabeth},
  year = {1976},
  journal = {European Journal of Social Psychology},
  volume = {6},
  number = {2},
  pages = {149--174},
  issn = {1099-0992},
  doi = {10.1002/ejsp.2420060202},
  urldate = {2024-08-09},
  copyright = {Copyright {\copyright} 1976 John Wiley \& Sons, Ltd},
  langid = {english}
}

@inproceedings{stauferSilencingRiskNot2024,
  title = {Silencing the {{Risk}}, {{Not}} the {{Whistle}}: {{A Semi-automated Text Sanitization Tool}} for {{Mitigating}} the {{Risk}} of {{Whistleblower Re-Identification}}},
  shorttitle = {Silencing the {{Risk}}, {{Not}} the {{Whistle}}},
  booktitle = {Proceedings of the 2024 {{ACM Conference}} on {{Fairness}}, {{Accountability}}, and {{Transparency}}},
  author = {Staufer, Dimitri and Pallas, Frank and Berendt, Bettina},
  year = {2024},
  month = jun,
  series = {{{FAccT}} '24},
  pages = {733--745},
  publisher = {Association for Computing Machinery},
  address = {New York, NY, USA},
  doi = {10.1145/3630106.3658936},
  urldate = {2024-07-21},
  isbn = {979-8-4007-0450-5}
}

\end{document}